\documentclass[a4paper,10pt]{article}

\usepackage{subfig}
\usepackage{graphicx}
\usepackage{cite}
\usepackage{bbm}
\usepackage{epsfig,amsmath,amssymb,verbatim,mathrsfs,array,layout,textcomp,amssymb,latexsym}

\newcommand{\beq}{\begin{eqnarray}}
\newcommand{\eeq}{\end{eqnarray}}
\newcommand{\be}{\begin{eqnarray}}
\newcommand{\ee}{\end{eqnarray}}

\def\beqa{\begin{eqnarray}}
\def\eeqa{\end{eqnarray}}
\def\bea{\begin{eqnarray}}
\def\eea{\end{eqnarray}}
\newcommand{\no}{\nonumber}
\newcommand{\bv}{\left(\begin{array}{c}}
\newcommand{\ev}{\end{array}\right)}
\newcommand{\bmtwo}{\left(\begin{array}{cc}}
\newcommand{\bmthree}{\left(\begin{array}{ccc}}
\newcommand{\emn}{\end{array}\right)}
\newcommand{\bmtwoc}{\left\{\begin{array}{cc}}
\newcommand{\bmthreec}{\left\{\begin{array}{ccc}}
\newcommand{\emnc}{\end{array}\right\}}
\newcommand{\ba}{\begin{array}}
\newcommand{\ea}{\end{array}}

\def\lsim{\mathrel{\rlap{\lower4pt\hbox{\hskip1pt$\sim$}}
     \raise1pt\hbox{$<$}}}         
\def\gsim{\mathrel{\rlap{\lower4pt\hbox{\hskip1pt$\sim$}}
     \raise1pt\hbox{$>$}}}         

\begin{document}

\begin{titlepage}

\vskip1.5cm
\begin{center}
  {\Large \bf Implications of the LHCb discovery of \\CP violation in charm decays}
\end{center}
\vskip0.2cm

\begin{center}
Avital Dery$^1$ and Yosef Nir$^2$\\
\end{center}
\vskip 8pt

\begin{center}
{ $^1$\it Department of Physics, LEPP, Cornell University ,Ithaca, NY 14853, USA}\\
{ $^2$\it Department of Particle Physics and Astrophysics,
Weizmann Institute of Science, Rehovot 76100, Israel} \vspace*{0.3cm}

{\tt   avital.dery@cornell.edu, yosef.nir@weizmann.ac.il}
\end{center}

\vglue 0.3truecm

\begin{abstract}
  \vskip 3pt \noindent
  The recent measurement of $\Delta A_{CP}$ by the LHCb collaboration requires an ${\cal O}(10)$ enhancement coming from hadronic physics in order to be explained within the SM. We examine to what extent can NP models explain $\Delta A_{CP}$ without such enhancements. We discuss the implications in terms of a low energy effective theory as well as in the context of several explicit NP models.

\end{abstract}

\end{titlepage}

\section{Introduction to $\Delta A_{CP}$}
%
The LHCb experiment has announced discovery of direct CP violation in singly Cabibbo suppressed $D$ decays \cite{Aaij:2019kcg},
\beqa
\Delta A_{CP}&\equiv&A_{CP}(K^+K^-)-A_{CP}(\pi^+\pi^-)\no\\
&=&(-1.54\pm0.29)\times10^{-3}.
\eeqa
Here
\beq
A_{CP}(f)\equiv\frac{\Gamma(D^0\to f)-\Gamma(\overline{D}{}^0\to f)}{\Gamma(D^0\to f)+\Gamma(\overline{D}{}^0\to f)}.
\eeq
In $\Delta A_{CP}$ effects of indirect CP violation approximately cancel out \cite{Grossman:2006jg}. (Due to different decay time acceptances between the $K^+K^-$ and $\pi^+\pi^-$ modes, a small residual effect of indirect CP violation remains.) Thus, $\Delta A_{CP}$ is a manifestation of CP violation in decay. The updated world average for the direct and indirect CP violating contributions to this asymmetry are \cite{Amhis:2019ckw}
\beq\label{eq:acpdir}
\Delta A_{CP}^{\rm dir}=(-1.64\pm0.28)\times10^{-3},\
\eeq
\beq
A_{CP}^{\rm ind}=(+0.28\pm0.26)\times10^{-3}.\
\eeq

The singly Cabibbo suppressed $D^0$ ($\overline{D}{}^0$) decay amplitudes $A_f$ ($\overline{A}_f$) to a final CP eigenstate $f$ can be written as \cite{Grossman:2006jg}
\beqa
A_f&=&A_f^Te^{i\phi_f^T}\left[1+r_f e^{i(\delta_f+\phi_f)}\right],\\
\overline{A}_f&=&\eta_{CP}A_f^Te^{-i\phi_f^T}\left[1+r_f e^{i(\delta_f-\phi_f)}\right],\no
\eeqa
where $\eta_{CP}=\pm1$ is the CP eigenvalue of $f$, the dominant singly Cabibbo suppressed ``tree" amplitude is denoted by $A_f^Te^{\pm i\phi_f^T}$, and $r_f$ parameterizes the relative magnitude of all subleading amplitudes (often called ``penguin" amplitudes), which carry different strong ($\delta_f$) and weak ($\phi_f$) phases. Then
\beq
A_{CP}^{\rm dir}(f)=-\frac{2r_f\sin\delta_f\sin\phi_f}{1+2r_f\cos\delta_f\cos\phi_f+r_f^2}.
\eeq

The Standard Model (SM) contribution to the individual asymmetries is CKM suppressed by a factor of
\beq\label{eq:ickm}
I_{\rm CKM}\equiv2{\cal I}m\left(\frac{V_{ub}V_{cb}^*}{V_{us}V_{cs}^*}\right)\approx 1.4\times10^{-3}.
\eeq
%
Naively, there is a further loop suppression by a factor of order $\alpha_s/\pi\sim0.1$.
One cannot exclude an enhancement factor of order 10 from hadronic physics \cite{Brod:2011re,Brod:2012ud,Grossman:2019xcj,Cheng:2019ggx}, in which case (\ref{eq:acpdir}) is accounted for by SM physics. Yet, it is not implausible that new physics (NP) dominates $\Delta A_{CP}$ \cite{Isidori:2011qw,Chala:2019fdb} (indeed, QCD-based LCSR calculations \cite{Khodjamirian:2017zdu} support the latter option.)

In the following we assume that hadronic factors do not significantly alter the magnitude of the relevant effects; Thus, NP is required to explain the measured $\Delta A_{CP}$. We analyze the implications of Eq.~(\ref{eq:acpdir}) on candidate models.
We phrase our findings in terms of which NP models can or cannot account for the measurement, assuming that the SM contribution is negligible. Relaxing this assumption, the same implications can be conservatively read as upper bounds on the NP model parameters.

In 2011, experimental evidence for $\Delta A_{CP}$~\cite{Aaij:2011in} prompted several related studies~\cite{Isidori:2011qw,Brod:2012ud,Hochberg:2011ru,Giudice:2012qq,Hiller:2012wf,Brod:2011re}. We provide an update to some of the relevant results, taking into account the recent discovery with a central value smaller by a factor of $\sim 4$ as well as all applicable existing bounds.

We begin with an effective field theory (EFT) analysis in Section~\ref{sec:EFT}. We follow with specific examples of models in which the measured $\Delta A_{CP}$ is explained: 2HDM in Section~\ref{sec:2HDM}, the MSSM in Section~\ref{sec:SUSY} and models with vector-like up-quarks in Section~\ref{sec:vectorlike}. We conclude in Section~\ref{sec:con}.

\section{Non-renormalizable operators}
\label{sec:EFT}
The relevant effects of new physics at a scale much higher than the electroweak breaking scale can be represented by the following effective Hamiltonian \cite{Isidori:2011qw}:
\beq\label{eq:Hdeltac1}
{\cal H}_{|\Delta c|=1}^{{\rm eff-NP}}=\frac{G_F}{\sqrt2}\sum_{i=1,2,5,6}\sum_q(C_i^qQ_i^q+C_i^{q\prime}Q_i^{q\prime})
+\frac{G_F}{\sqrt2}\sum_{i=7,8}(C_i^qQ_i+C_i^{\prime}Q_i^{\prime})+{\rm h.c.},
\eeq
where $q=\{d,s,b,u,c\}$, the list of operators includes
\beqa
Q_1^q&=&(\bar uq)_{V-A}(\bar qc)_{V-A},\,\,\,\,\,\, \quad\qquad Q_7=-\frac{e}{8\pi^2}m_c\bar u\sigma_{\mu\nu}(1+\gamma_5)F^{\mu\nu}c, \no\\
Q_2^q&=&(\bar u_\alpha q_\beta)_{V-A}(\bar q_\beta c_\alpha)_{V-A}, \qquad Q_8=-\frac{g_s}{8\pi^2}m_c\bar u\sigma_{\mu\nu}(1+\gamma_5)T^a G_a^{\mu\nu}c, \no\\
Q_5^q&=&(\bar uc)_{V-A}(\bar qq)_{V+A},\no\\
Q_6^q&=&(\bar u_\alpha c_\beta)_{V-A}(\bar q_\beta q_\alpha)_{V+A},
\eeqa
and the primed operators are related to the non-primed ones via $A\leftrightarrow -A$ and $\gamma_5\leftrightarrow-\gamma_5$.

The SM and NP contributions to $\Delta A_{CP}$ can be parameterized as
\be\label{eq:NPcont}
	\Delta A_{CP}\approx I_{\rm CKM} \frac{\alpha_s(m_c)}{\pi}{\cal I}m(\Delta R^{\rm SM}) +\frac{2}{|V_{us}V_{cs}|} \sum_i {\cal I}m(C_i^{\rm NP}){\cal I}m(\Delta R_i^{\rm NP}),
\ee
where $\Delta R^{\rm SM,NP} = R_K^{\rm SM,NP} +R_\pi^{\rm SM,NP} $, and $R_{K}^{{\rm SM,NP}}$ are the ratios of subleading amplitudes to the leading SM amplitude, after factorizing out the CKM dependence and the Wilson coefficient (the loop factor for $R_K^{\rm SM}$).
Thus the SM alone can explain the measured value of $\Delta A_{CP}$ for ${\cal I}m(\Delta R^{\rm SM})\approx 13$.
In the following we conversely adopt the naive expectation, ${\cal I}m(\Delta R^{\rm SM})\sim {\cal I}m(\Delta R^{\rm NP}) \sim 2$ (the factor of $2$ is inspired by the U-spin limit, in which $A^{\rm SM}_{K^+K^-}\approx -A^{\rm SM}_{\pi^+\pi^-}$.)
With this assumption, the measurement requires the existence of NP with a Wilson coefficient satisfying
\beq\label{eq:dacpcinp}
	{\cal I}m(C_i^{\rm NP})\sim\frac{\Delta A_{CP}}{18.2}\sim 9\times10^{-5},
\eeq
and the scale of NP can naively be estimated as $\lsim 37$ TeV.

\subsection{Constraints from $D^0-\overline{D}{}^0$ mixing}
The Hamiltonian of Eq.~(\ref{eq:Hdeltac1}) is related to the effective Hamiltonian relevant for $|\Delta c|=2$ transitions,
\be
{\cal H}^{\rm eff}_{|\Delta c|=2} = \frac{G_F}{\sqrt{2}}\left(\sum_{i=1}^5 C_i^{cu}Q_i^{cu}+ \sum_{i=1}^3 {C_i^{cu}}^\prime {Q_i^{cu}}^\prime \right),
\ee
where
\bea
Q_1^{cu} &=& (\bar u c)_{V-A}(\bar u c)_{V-A}  \qquad \qquad \quad Q_2^{cu} = (\bar u c)_{S-P}(\bar u c)_{S-P}\\ \no
Q_3^{cu} &=& (\bar u_\alpha c_\beta)_{S-P}(\bar u_\beta c_\alpha)_{S-P} \qquad \quad Q_4^{cu} = (\bar u c)_{S-P}(\bar u c)_{S+P}\\ \no
Q_5^{cu} &=& (\bar u_\alpha c_\beta)_{S-P}(\bar u_\beta c_\alpha)_{S+P}.
\eea
The contributions of ${\cal H}^{\rm eff}_{|\Delta c|= 2}$ to $D^0-\overline{D}{}^0$ mixing are computed using the following formula:
\be
\langle \bar D^0 |{\cal H}^{\rm eff}_{|\Delta c|= 2}|D^0 \rangle _i = \frac{G_F}{\sqrt{2}}\sum_{j=1}^5\sum_{r=1}^5 \left(b_j^{(r,i)} + \eta c_j^{(r,i)}	 \right)\eta^{a_j}\times C_i^{cu}(\mu) \langle \bar D^0 |Q_r^{cu}|D^0 \rangle,
\ee
where all relevant parameters and hadronic matrix elements are defined in Ref.~\cite{Bona:2007vi}.

Using the up-to-date 95\% C.L regions for the mixing parameters~\cite{Amhis:2019ckw},
\bea
x_{12} &\in & [0.22,0.63]\% \\ \no	
y_{12} &\in & [0.59,0.75]\%  \\ \no	
\phi_{12} &\in & [-2.5^o,1.8^o],
\eea
we obtain the following bounds:
\bea \label{eq:updatedDDbar}
{\rm Im}(C_1^{cu}) &\lsim & 1.6\times 10^{-9};\,\, \qquad {\rm Re}(C_1^{cu}) \lsim  3.6\times 10^{-8}, \\ \no
{\rm Im}(C_4^{cu}) &\lsim & 1.7\times 10^{-10}; \qquad {\rm Re}(C_4^{cu}) \lsim  4.0\times 10^{-9}, \\ \no
{\rm Im}(C_5^{cu}) &\lsim & 4.9\times 10^{-10}; \qquad {\rm Re}(C_5^{cu}) \lsim  1.1\times 10^{-8}.
\eea
Following Ref.~\cite{Isidori:2011qw}, we can relate the two sets of Wilson coefficients via
\bea\label{eq:deltac1rel}
C_1^{cu} &=& \delta C_1^{cu}+ \frac{g^2}{32\pi^2}\sum_q \lambda_q (C_2^q-C_1^q)\ln \frac{\mu^2}{m_W^2},\\ \no
C_4^{cu} &=& \delta C_4^{cu}- \frac{g^2}{16\pi^2}\sum_q \lambda_q {C_6^q}^\prime \ln \frac{\mu^2}{m_W^2},\\ \no
C_5^{cu} &=& \delta C_5	^{cu}- \frac{g^2}{16\pi^2}\sum_q \lambda_q {C_ 5^q}^\prime \ln \frac{\mu^2}{m_W^2}.
\eea
\begin{table}[!t]
	\caption{Upper bounds on CP violating $\Delta c=1$ operators from $D^0-\overline{D}{}^0$ mixing, at the hadronic charm scale $\mu \approx 2\,{\rm GeV}$.}
	\label{tab:dmixcon}
	\begin{center}
		\begin{tabular}{c|cc} \hline\hline
			\rule{0pt}{1.2em}%
			$f$ & $s-d$ & $8d$ \cr
			\hline \hline
			${\cal I}m(C_{1,2}^{(f)})$ & $3.6\times10^{-7}$ & $9.6\times10^{-4}$ \\
			
			${\cal I}m(C_{5}^{(f)\prime})$ & $5.6\times10^{-8}$ & $1.5\times10^{-4}$ \\
		
			${\cal I}m(C_{6}^{(f)\prime})$ & $2.0\times10^{-8}$ & $5.3\times10^{-5}$ \\		
			\hline\hline
		\end{tabular}
	\end{center}
\end{table}
We then change basis to $Q_i^{s-d} = Q_i^s-Q_i^d, Q_i^{8d} = Q_i^s+Q_i^d-2Q_i^b$,
and
take the counter-terms to zero to arrive at the bounds on the $\Delta c=1$ operators, presented in Table~\ref{tab:dmixcon}. We conclude that the operators $Q_{1,2}^{(s-d)}$, $Q_{5,6}^{(s-d)\prime}$ and $Q_{6}^{(8d)\prime}$ cannot account for $\Delta A_{CP}$.

\subsection{Constraints from $\epsilon^\prime/\epsilon$}
Following Ref.~\cite{Isidori:2011qw}, we use the master formula for $\epsilon^\prime/\epsilon$, evaluating the matrix elements induced by the $|\Delta s|=1$ operators at the large $N_c$ limit. The NP contribution is then given by
\be\label{eq:largeNcepsilon}
	\Big|\frac{\epsilon^\prime}{\epsilon}\Big|_{\rm NP} &\approx & 10^2\Big| {\cal I}m\Big[3.5 C_1^{(3/2)}+3.4C_2^{(3/2)}-1.7\rho^2C_5^{(3/2)}-5.2\rho^2C_6^{(3/2)}\\ \no
	&-& 0.04 C_1^{(1/2)} - 0.12C_2^{(1/2)} - 0.04\rho^2C_5^{(1/2)} + 0.11\rho^2C_6^{(1/2)}\Big]\Big|,
\ee
where $C_i^{(3/2)}= \frac{1}{2}(-C_i^{(s-d)}+C_i^{(c-u)}+C_i^{(8d)})+\frac{5}{4}C_i^{(b)}$, $C_i^{(1/2)} = \frac{1}{2}(C_i^{(s-d)}+C_i^{(c-u)}-C_i^{(8d)})+\frac{1}{4}C_i^{(b)}-C_i^{(0)}$, and $\rho = m_K/m_s$. Taking the conservative bound $|\epsilon^\prime/\epsilon|_{\rm NP} < |\epsilon^\prime/\epsilon|_{\rm exp}\approx 1.7\times 10^{-3}$, the imaginary parts of the $|\Delta s|=1$ Wilson coefficients are constrained. These are related to the $|\Delta c|=1$ coefficients of interest via
\be
	C_i^{q(ds)} = \delta C_i^{q(ds)} +C_i^q \frac{g^2}{32\pi^2}\ln\frac{\mu^2}{m_W^2}.
\ee

The resulting bounds on the $|\Delta c|=1$ Wilson coefficients are presented in Table~\ref{tab:epcon}. Comparing these bounds to Eq.~(\ref{eq:dacpcinp}), we conclude that the operators $Q_{5,6}^{(f)}$ with $f\in \{s-d,c-u,8d, b\}$ cannot account for $\Delta A_{CP}$.
\begin{table}[!ht]
	\caption{Upper bounds on CP violating $\Delta c=1$ operators from $|\epsilon^\prime/\epsilon|$, at the hadronic charm scale $\mu \approx  2\,{\rm GeV}$.}
	\label{tab:epcon}
	\begin{center}
		\begin{tabular}{c|cccccc} \hline\hline
			\rule{0pt}{1.2em}%
			$f$ & $s-d$ & $c-u$ & $8d$ & $b$ & $0$ \cr
			\hline \hline

			${\cal I}m(C_{1}^{(f)})$ & $4.8\times 10^{-4}$ & $4.9\times 10^{-4}$ & $4.8\times 10^{-4}$ & $1.9\times 10^{-4}$ & $2.1\times 10^{-2}$ \\
	
			${\cal I}m(C_{2}^{(f)})$ & $4.8\times 10^{-4}$ & $5.0\times 10^{-4}$ & $4.8\times 10^{-4}$ & $2.0\times 10^{-4}$ & $6.9\times 10^{-3}$ \\

			${\cal I}m(C_{5}^{(f)})$  & $3.6\times 10^{-5}$ & $3.4\times 10^{-5}$ & $3.6\times 10^{-5}$ & $1.4\times10^{-5}$ & $7.4 \times 10^{-4}$\\

			${\cal I}m(C_{6}^{(f)})$ & $1.1 \times 10^{-5}$ & $1.1 \times 10^{-5}$ & $1.1 \times 10^{-5}$ & $4.6\times 10^{-6}$  & $2.7\times 10^{-4}$ \\		
			\hline\hline
		\end{tabular}
	\end{center}
\end{table}

We note that the set of operators, $\{Q_{7,8}, Q_{7,8}^\prime, \forall f\, Q_{1,2}^{f\prime}, Q_{5,6}^{(c-u,b,0)\prime}\}$, are relevant to neither $D^0-\overline{D}{}^0$ mixing nor $|\epsilon^\prime/\epsilon|$, and therefore are unconstrained.
Table~\ref{tab:opList} summarizes which $\Delta c=1$ operators can contribute to $\Delta A_{CP}$ at a level comparable to the current measured value.

\begin{table}[!ht]
	\caption{Classification of new physics operators $Q_i$ according to whether upper bounds on ${\cal I}m(C_i^{\rm NP})$  from $D^0-\overline{D}{}^0$ mixing and $\epsilon^\prime/\epsilon$ are (i) much weaker than $9\times10^{-5}$ (``allowed"), (ii) of order $9\times10^{-5}$ (``marginal"), or (iii) much stronger than $9\times10^{-5}$ (``disfavored"). }
	\label{tab:opList}
	\begin{center}
		\begin{tabular}{p{4cm}p{4cm}p{4cm}} \hline
			\rule{0pt}{1.2em}%
			Allowed & Marginal & Disfavored \cr
			\hline \hline
			$Q_{7,8}, \,Q_{7,8}^\prime,$  & $Q_{5}^{(8d)\prime}$  & $Q_{1,2}^{(s-d)},$ \cr
			$\forall f\, Q_{1,2}^{f\prime}, \, Q_{5,6}^{(c-u,b,0)\prime}$,  & $Q_{6}^{(0)}$ & $Q_{5,6}^{(s-d)\prime},\,Q_{6}^{(8d)\prime},$ \cr
			$Q_{1,2}^{(c-u,8d,0)},\, Q_5^{(0)}$ & $Q_{1,2}^{(b)}$  & $Q_{5,6}^{(s-d,c-u,8d,b)},$\cr
			\hline
		\end{tabular}
	\end{center}
\end{table}
%


\section{2HDM}
\label{sec:2HDM}
As a first example of an explicit NP model that can account for the measurement of $\Delta A_{CP}$, we consider a two-Higgs-doublet model (2HDM), where a second scalar doublet,
\beq
	\Phi\sim (1,2)_{-1/2}=\left(\begin{array}{c} \phi^0 \\ \phi^- \\ \end{array} \right),
\eeq
is added to the SM. A contribution to $\Delta A_{CP}$ arises if $\phi^0$ couples to $u\bar u$ and $c\bar u$, generating both $D^0\to K^+K^-$ and $D^0\to \pi^+\pi^-$. Since all couplings besides $u\bar u$ and $c\bar u$ are irrelevant to this analysis, we take a conservative approach, considering minimal examples where $\Phi$ couples to $u_R$ and is aligned with a single down-type LH mass eigenstate. This allows us to evade tree-level scalar mediated FCNC in the down sector.
Assuming alignment with the quark doublet that has $b_L$ as its down-type quark, we have~\cite{Hochberg:2011ru}
\beq\label{eq:2hdmlag1}
{\cal L}_{\Phi}=-V(\Phi)+2\lambda\left[\phi^0\overline{U_{Li}} V_{ib}u_R+\phi^- \overline{b_L} u_R+{\rm h.c.}\right],
\eeq
where $U_{L1,2,3}=u_L,c_L,t_L$. Thus, the neutral scalar $\phi^0$ couples $u_R$ to $u_L$ and $c_L$: $\lambda V_{cb}\phi^0\bar c_L u_R+\lambda V_{ub}\phi_u^0\bar u_L u_R$. Integrating out the $\phi^0$ field, these couplings lead to the effective four-quark coupling.
\beq\label{eq:cuuuphi}
-\frac{8|\lambda|^2}{m_{\phi^0}^2}V_{ub}V_{cb}^*(\bar u_R c_L)(\bar u_L u_R) =\frac{|\lambda|^2}{m_{\phi^0}^2}V_{ub}V_{cb}^* Q_6^u.
\eeq
%
%
The contribution to $\Delta A_{CP}$, using Eq.~(\ref{eq:NPcont}), can be written as
\beq
	\Delta A_{CP}^{\phi} &\approx & \frac{2\sqrt{2}}{4|V_{us}V_{cs}|}\frac{G_0}{G_F}{\cal I}m(V_{ub}V_{cb}^*){\cal I}m(\Delta R^{\phi}) \\ \no
	&=& \frac{\sqrt{2}}{4}\frac{G_0}{G_F}I_{\rm CKM}\times{\cal I}m(\Delta R^{\phi})
\eeq
where
\beq
G_0\equiv 4|\lambda|^2/ m_{\phi^0}^2,
\eeq
%
and $I_{\rm CKM}$ is defined in Eq. (\ref{eq:ickm}).
What is needed then to account for (\ref{eq:acpdir}) is
%
\beq\label{eq:gzerogf}
	\frac{G_0}{G_F}\simeq \frac{3.3}{{\cal I}m(\Delta R^{\phi})}\ \Longrightarrow\ {\cal I}m(\Delta R^{\phi}) G_0\simeq\frac{1}{(160\ {\rm GeV})^2}.
\eeq
Thus, for ${\cal I}m(\Delta R^{\phi})\in\{0.2-2\}$, we need $G_0^{-1/2}=m_{\phi^0}/(2|\lambda|)\in\{70,230\}$ GeV.

\subsection{Constraints from $D^0-\overline{D}{}^0$ mixing}
The scalar exchange contributes to $D^0-\overline{D}{}^0$ mixing via box diagrams. Requiring that this contribution is not larger than the experimental constraints from $\Delta m_D$ gives \cite{Hochberg:2011ru}
\beq
\frac{|\lambda|^4}{32\pi^2}\left(\frac{100\ {\rm GeV}}{m_{\phi_u^0}}\right)^2(V_{ub}V_{cb}^*)^2<7\times10^{-9},
\eeq
or, equivalently
\beq\label{eq:ddbarbound}
\frac{|\lambda|^2 G_0}{G_F}<3\times10^3,
\eeq
so, taking into account (\ref{eq:gzerogf}), the new contribution is negligible, allowing for the required $G_0/G_F$ to explain $\Delta A_{CP}$.

\subsection{Constraints from $\epsilon^\prime/\epsilon$}
The same Yukawa couplings of $\phi^0$ that contribute to direct CP violation in $D$ decays, contribute unavoidably also to direct CP violation in $K$ decays. The former effect comes at tree level and modifies $\Delta A_{CP}$. The latter effect comes via box diagrams, involving $\phi^0$ and a $W$-boson, and modifies $\epsilon^\prime/\epsilon$. Upon integration out of $\phi^0$ and $W$, we obtain the following effective four-quark coupling:
\beq\label{eq:effsuud}
	\frac{\sqrt2|\lambda|^2 G_F}{\pi^2}\Big(f_{uu}(x_\phi)-2f_{ut}(x_\phi)+f_{tt}(x_\phi)\Big)V_{td}^*V_{ts}|V_{tb}|^2(\bar d_L u_R)(\bar u_R s_L),
\eeq
where $x_\phi\equiv m_{\phi^0}^2/m_W^2$, and the loop function is given by
\be
	f_{ij}(x_\phi) = \frac{x_i^2 \log x_i}{(1-x_i)(x_j-x_i)(x_\phi-x_i)}+\frac{x_j^2 \log x_j}{(1-x_j)(x_i-x_j)(x_\phi-x_j)}+\frac{x_\phi^2 \log x_\phi}{(1-x_\phi)(x_i-x_\phi)(x_j-x_\phi)}.
\ee
Using the relation $(\bar d_L u_R)(\bar u_R s_L) = -\frac{1}{8}(\bar d_\alpha s_\beta)_{V-A}(\bar u_\beta u_\alpha)_{V+A} = -\frac{1}{8}Q_6^{u(ds)}$, we read off the corresponding Wilson coefficient,
\be
	C_6^{u(ds)} = -\frac{|\lambda|^2 }{4\pi^2}\Big(f_{uu}(x_\phi)-2f_{ut}(x_\phi)+f_{tt}(x_\phi)\Big)V_{td}^*V_{ts}|V_{tb}|^2.
\ee
Following Ref.~\cite{Kagan:1999iq}, we use
 \beq
 {\cal R}e\left(\frac{\epsilon^\prime}{\epsilon}\right) = -\frac{\omega}{\sqrt{2}|\epsilon|}\left(\frac{{\cal I}m(A_0)}{{\cal R}e(A_0)}-\frac{{\cal I}m(A_2)}{{\cal R}e(A_2)}\right),
 \eeq
and
 \beq
 \frac{{\cal I}m A_2^\phi}{{\cal R}e A_2}\approx\frac32\frac{m_K^2}{m_s^2(m_c)-m_d^2(m_c)}
 \frac{{\cal I}m[\Delta C_6(m_c)+\frac13 \Delta C_5(m_c)]B_8^{(2)}(m_c)}{0.363|V_{us}^*V_{ud}|},
 \eeq
 where $\Delta C_i = C_i^u-C_i^d$.
 At the matching scale, our model generates $\Delta C_6(m_{\phi^0}) = C_6^u(m_{\phi^0})$, and $\Delta C_5(m_{\phi^0}) = 0$.
 Taking the conservative bound ${\cal R}e(\epsilon^\prime/\epsilon)^\phi < {\cal R}e(\epsilon^\prime/\epsilon)^{\rm Exp} \approx 1.66\times 10^{-3}$, we reach the constraint
 \be
 `C_6^{u(ds)}(m_{\phi^0}) < 2.23 \times 10^{-7}.
 \ee
 \begin{figure}[!t]
 	\centering
 	\includegraphics[width=5.5in]{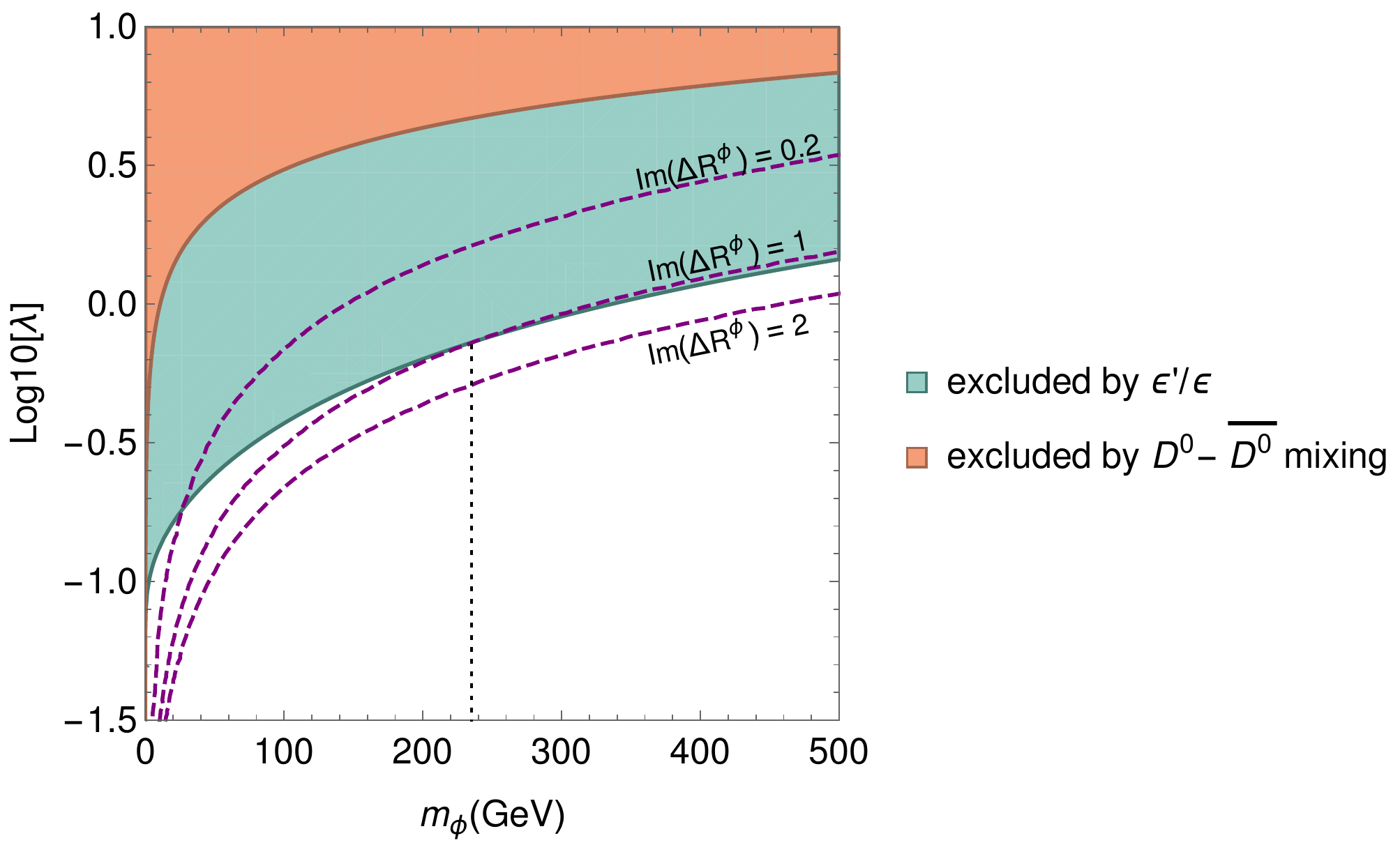}
 	\caption{Excluded regions in parameter space due to $D^0-\overline{D}{}^0$ mixing and $\epsilon^\prime/\epsilon$ constraints. The dashed lines depict the curves for which $\Delta A_{CP}$ is explained for ${\cal I}m(\Delta R^\phi) = 0.2,1,2$. The dotted vertical line marks the intersection, at $m_\phi\simeq 235\,{\rm GeV}$.}
 	\label{fig:scalar}
 \end{figure}
Figure~\ref{fig:scalar} presents the various constraints together with curves for which Eq.~(\ref{eq:gzerogf}) is satisfied with three representative values taken for ${\cal I}m(\Delta R^{\phi})$.
We conclude that $\Delta A_{CP}$ can be explained within this model, depending on the value of ${\cal I}m(\Delta R^{\phi})$. For ${\cal I}m(\Delta R^{\phi}) \approx 1$, the mass of the neutral scalar is bounded to be $m_\phi\lsim 235 \,{\rm GeV}$, while for ${\cal I}m(\Delta R^{\phi}) \approx 0.2$ it is bounded to be very light, and subject to further constraints. For ${\cal I}m(\Delta R^{\phi}) \gtrsim 1.5$, the mass is unconstrained.

We note the following points:
\begin{itemize}
	\item Two alternative choices for the Yukawa matrices such that only one down-type mass eigenstate is involved exist, with $\Phi$ aligned with the doublet containing either $d_L$ or $s_L$. These suffer from large contributions to $D^0-\overline{D}{}^0$ mixing, and therefore cannot account for $\Delta A_{CP}$.
	
	\item It may seem surprising that this model can account $\Delta A_{CP}$ even though it contributes via the operator $Q_6^{c-u}$, disfavored by the EFT analysis. This is explained by the existence of additional contributions within this model to $\epsilon/\epsilon^\prime$, which interfere destructively. These are not taken into account in the EFT approach. Therefore this model evades the EFT conclusions regardless of the mass scale of the new scalars.
	
	\item We note that mid-range masses for the charged scalar ($450\,{\rm GeV} \lsim m_{\phi^-}\lsim $ a few TeV) are constrained by LHC dijet searches~\cite{Aad:2014aqa,Aaboud:2018fzt,Sirunyan:2018xlo}. These would result in a further constraint in the $(|\lambda|,m_{\phi^0})$ plane, depending on the mass splitting between the neutral and charged scalars. Charged scalar masses below $450\,{\rm GeV}$ or above a few TeV are not constrained by these bounds. 
\end{itemize}

\section{MSSM}
\label{sec:SUSY}
As a second example for candidate NP models to explain the measurement of $\Delta A_{CP}$, we consider the MSSM. The dominant supersymmetric contribution to $\Delta A_{CP}$ is likely to come from loops involving gluinos and up-squarks. These contribute to the chromomagnetic operators $Q_8$ and $Q_8^\prime$, which are very weakly constrained by $D^0-\overline{D}{}^0$ mixing and $\epsilon/\epsilon^\prime$. The dominant source of CP violation is likely to be the chirality-changing and flavor-changing mass-squared insertion \cite{Giudice:2012qq},
\beq
\delta_{LR}\equiv(\delta^u_{LR})_{12}=\frac{(\tilde M^{2u}_{LR})_{12}}{\tilde m^2},
\eeq
where $\tilde m^2$ is the average up-squark mass, and $\tilde M^{2u}_{LR}$ is the left-right block in the $6\times6$ up-squark mass-squared matrix. In the approximation that only two squark generations are involved, we can express this parameter in terms of the supersymmetric mixing angles, $(K^u_{L,R})_{ij}$ and the mass-squared splitting between the squarks, $\Delta\tilde m^2_{ij}$:
\beq
\delta_{LR}=\frac{\Delta\tilde m^2_{q_{L1}q_{R2}}}{\tilde m^2}(K^u_L)_{12}(K^u_R)_{22}.
\eeq
One can estimate the supersymmetric contribution as \cite{Giudice:2012qq}%
\beq\label{eq:dacpsusy}
\Delta A_{CP}=1.5\times10^{-3}\frac{{\cal I}m(\delta_{LR})}{2.5\times10^{-4}}\frac{1\ {\rm TeV}}{\tilde m}\times {\cal I}m(\Delta R^{\rm SUSY}).
\eeq
Thus in order to explain $\Delta A_{CP}$ we require
\beq\label{eq:deltareq}
	{\cal I}m(\delta_{LR}) \approx 2.5\times10^{-4}\frac{\tilde m}{1\ {\rm TeV}}{\cal I}m(\Delta R^{\rm SUSY})^{-1}.
\eeq
In MFV models \cite{Hiller:2012wf},
\beq
\delta_{LR}\propto\frac{m_c}{\tilde m}(y_s^2V_{us}V_{cs}^*+y_b^2 V_{ub}V_{cb}^*)\lsim10^{-7},
\eeq
and the contribution is negligible. In Froggatt-Nielsen (FN) models \cite{Nir:2002ah,Hiller:2012wf},
\beq\label{eq:dlrfn}
\delta_{LR}\sim\frac{\tilde a}{\tilde m}\frac{m_c|V_{us}|}{\tilde m}\sim3\times10^{-4}\frac{\tilde a}{\tilde m}\frac{1\ {\rm TeV}}{\tilde m},
\eeq
where $\tilde a$ is the typical scale of the trilinear scalar coupling. When comparing Eq. (\ref{eq:dlrfn}) to Eq. (\ref{eq:deltareq}), it seems that FN-SUSY models are plausible candidates to account for $\Delta A_{CP}$. One has to take into account, however, the FN relations with other entries of the squark mass-squared matrices, and, in particular,
\beq
\frac{{\cal I}m(\delta_{LR}^u)_{12}}{{\cal I}m(\delta_{LR}^q)_{11}}\sim\frac{m_c|V_{us}|}{m_q},\ \ \ (q=u,d).
\eeq
Assuming phases of order one (which we do to explain $\Delta A_{CP}$), the flavor-diagonal parameters are bounded by the EDM constraints. The resulting bounds are \cite{Hiller:2012wf}
\beqa\label{eq:EDMconst}
(\delta_{LR}^u)_{12}&\lsim&3\times10^{-4}\frac{\tilde m}{{\rm TeV}}\ \ \ (\text{from } (\delta_{LR}^u)_{11}),\no\\
(\delta_{LR}^u)_{12}&\lsim&8\times10^{-5}\frac{\tilde m}{{\rm TeV}}\ \ \ (\text{from } (\delta_{LR}^d)_{11}).
\eeqa
Comparing to Eq.~(\ref{eq:dacpsusy}), we see that within FN, ${\cal I}m(\Delta R^{\rm SUSY})\gtrsim 3$ is required in order to explain $\Delta A_{CP}$. In more elaborate flavor schemes (as in, for example, Ref.~\cite{Calibbi:2013mka}) it is possible that Eq.~(\ref{eq:deltareq}) is satisfied for ${\cal I}m(\Delta R^{\rm SUSY}) \approx 2$.

\section{Vector-like quarks}
\label{sec:vectorlike}
A third example for a model that may explain $\Delta A_{CP}$ is a model exhibiting flavor changing $Z$ couplings.
Models with extra non-sequential quarks generally induce such flavor changing couplings for the $Z$ boson. For example, the addition of vector-like up quarks in the $(3,1,+2/3)\bigoplus(\bar 3,1,-2/3)$ representation induces flavor changing $Z$ couplings of the form~\cite{Grossman:2006jg}
\beq
-{\mathcal L}_Z = \frac{gU_{ij}^u}{2\cos\theta_W}\overline{u_L}_i\gamma_\mu {u_L}_jZ^\mu+{\rm h.c.}
\eeq
The relevant coupling for $\Delta A_{CP}$ is $U_{cu}^u$, which also contributes at tree level to $\Delta m_D$, and at loop level to $\epsilon^\prime/\epsilon$.
\subsection{Constraint from $D^0-\overline{D}{}^0$ mixing}
The constraint from $\Delta m_D$ can be calculated using the effective operators of Ref.~\cite{Isidori:2011qw}. The relevant $\Delta c = 2$ operator is $(\bar u_L \gamma^\mu c_L)^2 = \frac{1}{4}Q_1^{cu}$.
Using Eq.~(\ref{eq:updatedDDbar}) for the current bound on ${\rm Re}(C_{1}^{cu})$, we arrive at
\beq\label{eq:UcuDeltambound}
	|U_{cu}^u|\lsim 2.8\times 10^{-4}.
\eeq
%
\subsection{Constraints from $\epsilon^\prime/\epsilon$}
A contribution to $\epsilon^\prime/\epsilon$ arises via a $W$-loop, inducing the operators $Q_{1,5}^{u(ds)} = (\bar u u)_{V\mp A}(\bar s d)_{V-A}$. We calculate the relevant Wilson coefficients and arrive at
\beq
	C_1^{u(ds)} &=& \frac{(3-4s_W^2) U_{cu}}{96\pi^2} \approx U_{cu}\cdot 2.2\times 10^{-3}, \\ \no
	C_5^{u(ds)} &=& \frac{s_W^2 U_{cu}V_{cs}V_{ud}^*}{24\pi^2} \approx U_{cu} \cdot 9.3\times 10^{-4 }.
\eeq
Using Eq.~(\ref{eq:largeNcepsilon}), the constraint on these coefficients is given by
\beq
	{\cal I}m(C_1^{(c-u)(ds)}) &\lsim & 9.8\times 10^{-6};\\ \no
	{\cal I}m(C_5^{(c-u)(ds)}) &\lsim & 7.1\times 10^{-7},
\eeq
The constraint on $C_5^{(c-u)(ds)}$ is more stringent, implying
 \be\label{eq:Ucuepsilonbound}
 	{\cal I}m(U_{cu})\lsim 7.6\times 10^{-4}.
 \ee

$\Delta A_{CP}$ arises in this model through the tree level annihilation diagram $\bar c u\to \bar u u $, which contributes to the $\Delta c=1$ four quark operators,
\beqa
(\bar u_L \gamma_\mu c_L)(\bar u_L \gamma^\mu u_L) &=& \frac{1}{4}Q_1^u, \\ \no
(\bar u_L \gamma_\mu c_L)(\bar u_R \gamma^\mu u_R) &=& \frac{1}{4}Q_5^u.
\eeqa
The coefficients of these operators in this model are given by
\beqa
C_1^u &=&  U_{cu}^u \left(\frac{1}{2}-\frac{2}{3}\sin^2\theta_W\right), \\ \no
C_5^u &=& U_{cu}^u \frac{2}{3}\sin^2\theta_W.
\eeqa
Using Eq.~(\ref{eq:NPcont}), the contribution to $\Delta A_{CP}$ can be written as
\beq
\Delta A_{CP} \approx \frac{2}{|V_{us}V_{cs}|}\left({\cal I}m(C_1^u){\cal I}m(\Delta R_1^Z)+ {\cal I}m(C_5^u){\cal I}m(\Delta R_5^Z)\right). 
\eeq
when we have taken ${\cal I}m(\Delta R_1^Z)\approx {\cal I}m(\Delta R_5^Z)\equiv {\cal I}m(\Delta R^Z)$.
Thus in order to explain the measurement we require
\beq
	{\cal I}m(U_{cu}^u)\approx 1.84\times 10^{-4} \left(\frac{2}{{\cal I}m(\Delta R^Z)}\right),
\eeq
which (under the assumption of ${\cal I}m(\Delta R^Z)\approx 2$) is allowed by Eqs.~(\ref{eq:UcuDeltambound},\ref{eq:Ucuepsilonbound}).

We note that this model is viable despite the fact that it induces the EFT-disfavored operator $Q_5^{(c-u)}$ (see Table~\ref{tab:opList}), as its contibution to $\Delta A_{CP}$ is subleading to that of the operator $Q_1^{(c-u)}$.

\section{Discussion}
\label{sec:con}
We have addressed the question of how easily can the new measurement of $\Delta A_{CP}$ be explained using benchmark NP models. We have followed the assumption that no significant hadronic enhancements are present, and derived the constraints coming mainly from measurements of $D^0-\overline{D}{}^0$ mixing and $\epsilon^\prime/\epsilon$. We find that non-generic though still simple NP models can account for the measured asymmetry.

Three candidate NP models were discussed -- 2HDM, MSSM and vector-like up-quarks. Our assumption of no significant hadronic enhancements is implemented by allowing at most ${\cal I}m(\Delta R^{\rm SM,NP})\approx 2$, in our Eq.~(\ref{eq:NPcont}).
We find that:
\begin{itemize}

	\item Both a 2HDM where scalar $(c \bar u),(u \bar u)$ couplings are present and models with vector-like up-quarks inducing $(c\bar u)$ $Z$ couplings can account for the measured asymmetry.
	
	\item The MSSM combined with flavor frameworks (MFV, FN) is unable to produce the desired contribution (FN requires ${\cal I}m(\Delta R^{\rm FN})\gtrsim 3$). The MSSM with a generic flavor structure is unconstrained.
	
\end{itemize}

Ref.~\cite{Grossman:2019xcj} studied the scenario where the SM accounts for $\Delta A_{CP}$ with mild $SU(3)$ breaking effects but a strong enhancement of $\Delta U=0$ transitions. They obtain two predictions: $U$-spin invariant strong phases should be large, and $A_{CP}(K^+K^-)\approx-A_{CP}(\pi^+\pi^-)$. Interestingly, in all three models that we analyzed the new physics operators that account for $\Delta A_{CP}$ do not introduce new sources of $U$-spin breaking, and thus the latter prediction does not favor the SM over these models.

In all three specific new physics models, the flavor structure is not in the minimal flavor violation class, and in fact it is non-generic. Thus, it is difficult to make definite predictions for the modification of other flavor changing and/or CP violating processes. Yet, it is unlikely that the only significant modification would be to singly Cabibbo suppressed charm decays. This situation motivates a broad flavor precision program, such as in the LHCb and BELLE-II experiments.

Of course, a direct search for the new degrees of freedom required by the various models is also well motivated. The upper bound on the scale of new physics is model dependent, and varies from few tens of TeV in the low energy EFT, to hundreds of GeV in the 2HDM.

\subsection*{Acknowledgements}
We thank Daniel Egana-Ugrinovic for helpful comments on the manuscript, and Daniel Aloni for useful discussions. YN is the Amos de-Shalit chair of theoretical physics, and is supported by grants from the Israel Science Foundation (grant number 394/16), the United States-Israel Binational Science Foundation (BSF), Jerusalem, Israel (grant number 2014230), and the I-CORE program of the Planning and Budgeting Committee and the Israel Science Foundation (grant number 1937/12).



\begin{thebibliography}{99}
%

\bibitem{Aaij:2019kcg}
R.~Aaij {\it et al.} [LHCb Collaboration],
Phys.\ Rev.\ Lett.\  {\bf 122}, no. 21, 211803 (2019)
[arXiv:1903.08726 [hep-ex]].

\bibitem{Grossman:2006jg}
Y.~Grossman, A.~L.~Kagan and Y.~Nir,
Phys.\ Rev.\ D {\bf 75}, 036008 (2007)
[hep-ph/0609178].


\bibitem{Amhis:2019ckw} 
  Y.~S.~Amhis {\it et al.} [Heavy Flavor Averaging Group],
  arXiv:1909.12524 [hep-ex].
  


\bibitem{Brod:2011re}
J.~Brod, A.~L.~Kagan and J.~Zupan,
Phys.\ Rev.\ D {\bf 86}, 014023 (2012)
[arXiv:1111.5000 [hep-ph]].

\bibitem{Brod:2012ud}
J.~Brod, Y.~Grossman, A.~L.~Kagan and J.~Zupan,
JHEP {\bf 1210}, 161 (2012)
[arXiv:1203.6659 [hep-ph]].

\bibitem{Grossman:2019xcj}
Y.~Grossman and S.~Schacht,
JHEP {\bf 1907}, 020 (2019)
[arXiv:1903.10952 [hep-ph]].

\bibitem{Cheng:2019ggx}
  H.~Y.~Cheng and C.~W.~Chiang,
  arXiv:1909.03063 [hep-ph].

\bibitem{Isidori:2011qw}
G.~Isidori, J.~F.~Kamenik, Z.~Ligeti and G.~Perez,
Phys.\ Lett.\ B {\bf 711}, 46 (2012)
[arXiv:1111.4987 [hep-ph]].

\bibitem{Chala:2019fdb}
M.~Chala, A.~Lenz, A.~V.~Rusov and J.~Scholtz,
 JHEP {\bf 1907}, 161 (2019)
[arXiv:1903.10490 [hep-ph]].

\bibitem{Khodjamirian:2017zdu}
A.~Khodjamirian and A.~A.~Petrov,
Phys.\ Lett.\ B {\bf 774}, 235 (2017)
[arXiv:1706.07780 [hep-ph]].


\bibitem{Aaij:2011in}
R.~Aaij {\it et al.} [LHCb Collaboration],
Phys.\ Rev.\ Lett.\  {\bf 108}, 111602 (2012)
[arXiv:1112.0938 [hep-ex]].


\bibitem{Hochberg:2011ru}
Y.~Hochberg and Y.~Nir,
Phys.\ Rev.\ Lett.\  {\bf 108}, 261601 (2012)
[arXiv:1112.5268 [hep-ph]].




\bibitem{Giudice:2012qq}
G.~F.~Giudice, G.~Isidori and P.~Paradisi,
JHEP {\bf 1204}, 060 (2012)
[arXiv:1201.6204 [hep-ph]].

\bibitem{Hiller:2012wf}
G.~Hiller, Y.~Hochberg and Y.~Nir,
Phys.\ Rev.\ D {\bf 85}, 116008 (2012)
[arXiv:1204.1046 [hep-ph]].


\bibitem{Bona:2007vi}
M.~Bona {\it et al.} [UTfit Collaboration],
JHEP {\bf 0803}, 049 (2008)
[arXiv:0707.0636 [hep-ph]].




\bibitem{Kagan:1999iq}
A.~L.~Kagan and M.~Neubert,
Phys.\ Rev.\ Lett.\  {\bf 83}, 4929 (1999)
[hep-ph/9908404].


\bibitem{Aad:2014aqa} 
  G.~Aad {\it et al.} [ATLAS Collaboration],
  Phys.\ Rev.\ D {\bf 91}, no. 5, 052007 (2015)
  [arXiv:1407.1376 [hep-ex]].
  
\bibitem{Aaboud:2018fzt} 
  M.~Aaboud {\it et al.} [ATLAS Collaboration],
  Phys.\ Rev.\ Lett.\  {\bf 121}, no. 8, 081801 (2018)
  [arXiv:1804.03496 [hep-ex]].
  
\bibitem{Sirunyan:2018xlo} 
  A.~M.~Sirunyan {\it et al.} [CMS Collaboration],
  JHEP {\bf 1808}, 130 (2018)
  [arXiv:1806.00843 [hep-ex]].

  
\bibitem{Nir:2002ah}
Y.~Nir and G.~Raz,
Phys.\ Rev.\ D {\bf 66}, 035007 (2002)
[hep-ph/0206064].


\bibitem{Calibbi:2013mka} 
  L.~Calibbi, P.~Paradisi and R.~Ziegler,
  JHEP {\bf 1306}, 052 (2013)
  [arXiv:1304.1453 [hep-ph]].


	




  




  


\end{thebibliography}
\end{document}